\documentclass[superscriptaddress, nobibnotes, aps, preprint, notitlepage]{revtex4-1}

\usepackage{booktabs}
\usepackage{graphicx}
\usepackage{natbib}
\usepackage{amsmath}%
\usepackage{amsfonts}%
\usepackage{amssymb}%
\usepackage{array}

\usepackage{checkend}	
\usepackage{xspace}
\usepackage{color}
\definecolor{greytext}{gray}{0.5}
\usepackage[caption=off]{subfig}

\usepackage[unicode=true,
  linktocpage,
  linkbordercolor={0.5 0.5 1},
  citebordercolor={0.5 1 0.5},
  linkcolor=blue]{hyperref}

\newcommand{\YBCO}[1]{YBa\ensuremath{_2}Cu\ensuremath{_3}O\ensuremath{_{6{.#1}}}\xspace}
\newcommand{\YBCOd}{YBa\ensuremath{_2}Cu\ensuremath{_3}O\ensuremath{_{6+\mathrm{\delta}}}\xspace}

\newcommand{\caxis}{\ensuremath{\hat{\mathrm{c}}}-axis\xspace}
\newcommand{\aaxis}{\ensuremath{\hat{\mathrm{a}}}-axis\xspace}
\newcommand{\baxis}{\ensuremath{\hat{\mathrm{b}}}-axis\xspace}
\newcommand{\abplane}{\ensuremath{\hat{\mathrm{a}}}-\ensuremath{\hat{\mathrm{b}}}-plane\xspace}
\newcommand{\degr}[1]{\ensuremath{#1^{\circ}}\xspace}

\newcommand{\kF}{\ensuremath{k_\mathrm{F}}\xspace}

\newcommand{\highTc}{high-\ensuremath{\mathrm{T}_\mathrm{c}}\xspace}

\newcommand{\Cfour}{\ensuremath{\mathrm{C}_4}\xspace}
\newcommand{\Ctwo}{\ensuremath{\mathrm{C}_2}\xspace}
\newcommand{\rhozz}{\ensuremath{\rho_{\mathrm{zz}}}\xspace}

\begin{document}

\title{Broken rotational symmetry on the Fermi surface of a high-T$_c$ superconductor}

\let\clearpage\relax
\author{B.~J.~Ramshaw}
\affiliation{Los Alamos National Labs, Los Alamos, NM, United States of America}
\email{Corresponding author: bradramshaw@gmail.com}
\author{N.~Harrison}
\affiliation{Los Alamos National Labs, Los Alamos, NM, United States of America}
\author{S.~E.~Sebastian}
\affiliation{Cavendish Laboratory, Cambridge University, Cambridge CB3 OHE, UK}
\author{S.~Ghannadzadeh}
\affiliation{High Field Magnet Laboratory, Radboud University, 6525 ED Nijmegen, The Netherlands}
\author{K.~A.~Modic}
\affiliation{Los Alamos National Labs, Los Alamos, NM, United States of America}
\affiliation{Present address: Max-Planck-Institute for Chemical Physics of Solids, Noethnitzer Strasse 40, D-01187, Dresden, Germany}
\author{D.~A.~Bonn}
\affiliation{Unviersity of British Columbia, Vancouver, Canada}
\author{W.~N.~Hardy}
\affiliation{Unviersity of British Columbia, Vancouver, Canada}
\author{Ruixing Liang}
\affiliation{Unviersity of British Columbia, Vancouver, Canada}
\author{P.~A.~Goddard}
\affiliation{Department of Physics, University of Warwick, Gibbet Hill Road, Coventry, CV4 7AL, U.K.}

\date{July 14, 2016}

\maketitle

\textbf{Broken fourfold rotational (${\rm C}_4$) symmetry is observed in the experimental properties of several classes of unconventional superconductors. It has been proposed that this symmetry breaking is important for superconducting pairing in these materials, but in the high superconducting transition temperature (\highTc) cuprates this broken symmetry has never been observed on the Fermi surface. We have measured a pronounced anisotropy in the angle dependence of the interlayer magnetoresistance of the underdoped \highTc superconductor \YBCO{58}, directly revealing broken \Cfour symmetry on the Fermi surface. Moreover, we demonstrate that this Fermi surface has \Ctwo symmetry of the type produced by a uniaxial or anisotropic density-wave phase. This establishes the central role of \Cfour symmetry breaking in the Fermi surface reconstruction of \YBCOd, and suggests a striking degree of universality among unconventional superconductors.
}

Broken \Cfour symmetry is observed in a number of experiments on unconventional superconductors, including transport \cite{Borzi:2007,Ando:2002,Daou:2010,Chu:2010}, NMR \cite{Wu:2011,Wu:2015}, neutron scattering \cite{Stock:2004,Hinkov:2008}, X-ray scattering \cite{Blanco:2014,Comin:2015,Chang:2015} and scanning tunnelling microscopy~\cite{Howald:2003,Kohsaka:2007}. In the iron-based superconductors broken ${\rm C}_4$ symmetry is observed directly on the Fermi surface \cite{Nakayama:2014}, and has been taken as an indication that this broken symmetry drives the high $\mathrm{T}_{\mathrm{c}}$s \cite{Fernandes:2014}. The question of whether the same type of symmetry breaking is relevant in the copper-oxide \highTc superconductors \cite{Kivelson:1998} has been left open because it has never been observed on the Fermi surface \cite{Sebastian:2014,Hossain:2008}. Without a link to the Fermi surface, it is hard to make a compelling argument that the experimental observations of broken ${\rm C}_4$ symmetry are relevant to the phenomenon of \highTc.

The Fermi surface of a metal is constrained by the symmetry of its electronic environment. A precise determination of Fermi surface geometry therefore provides information about symmetry-breaking states of matter that feed back into the electronic structure. In the underdoped \highTc cuprates it is now well established that a charge-density wave (CDW) competes with superconductivity, but a direct experimental connection is still missing between the CDW and the Fermi surface. QO (quantum oscillation) measurements, performed in the high-field state, have been instrumental in determining the presence of a small electron Fermi surface in the underdoped side of the phase diagram \cite{Doiron-Leyraud:2007,LeBoeuf:2007,Singleton:2010,Sebastian:2014,Ramshaw:2015b}. While QOs provide precise information about the area of the Fermi surface, they are relatively insensitive to its overall shape in 2D metals (e.g. circle vs. square). Here we present the results of a complementary technique, angle-dependent magnetoresistance (AMR), that determines the geometry and symmetry of the Fermi surface. We identify the shape of the Fermi surface as consistent with proposed CDW reconstruction scenarios \cite{Harrison:2011,Harrison:2012,Sebastian:2014,Maharaj:2014,Allais:2015,Briffa:2015}, and we find that this reconstruction is anisotropic in nature.  

Since the 1930s it has been known that a change in resistance with an applied magnetic field---magnetoresistance---provides geometric information about a metallic Fermi surface \cite{Jones:1934}, and early magnetoresistance experiments were instrumental in developing the modern quantum theory of metals \cite{Kapitza:1929}. Real metals have non-spherical Fermi surfaces, thus magnetoresistance can be a strong function of the angle between the magnetic field and the crystal axes, giving rise to AMR. The signatures of AMR are particularly strong for quasi-2D Fermi surfaces, and thus this technique is well suited for determining the Fermi surface geometry of the \highTc cuprates \cite{Singleton:2000,Bergemann:2003, Hussey:2003, Analytis:2007, Kartsovnik:2004}. To map the Fermi surface geometry of \YBCO{58} we performed interlayer (\rhozz) magnetoresistance measurements in a fixed magnetic field of 45~T. The field-angle dependence of \rhozz was obtained by rotating the sample in-situ, sweeping the polar angle $\theta$ between the magnetic field and the crystalline $\hat{c}$-axis for several values of the azimuthal angle $\phi$ (see the inset of Fig. 1a for angle definitions). Our primary observation is twofold anisotropy of the AMR as a function of $\phi$---immediately apparent in Fig. 1---with the AMR increasing much more rapidly on rotating the field towards the crystalline \aaxis ($\phi=$~0$^\circ$) than towards the \baxis ($\phi=$~90$^\circ$). As we demonstrate below, this directly indicates that Fermi surface has strongly broken \Cfour symmetry. 

\begin{figure}
\begin{minipage}[]{.55\columnwidth} 
\subfloat{
\includegraphics[width=1\columnwidth]{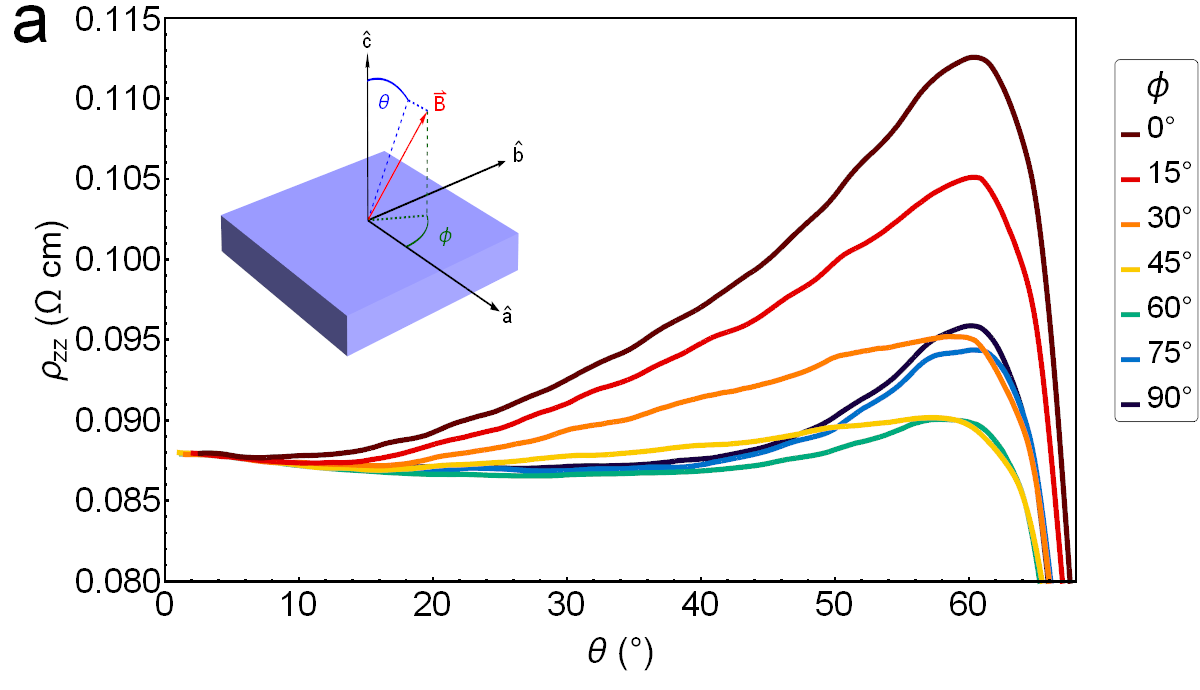}
\label{fig:osc1D}
}
\end{minipage}
\begin{minipage}[]{.44\columnwidth} 
\subfloat{
\includegraphics[width=1\columnwidth, clip=true]{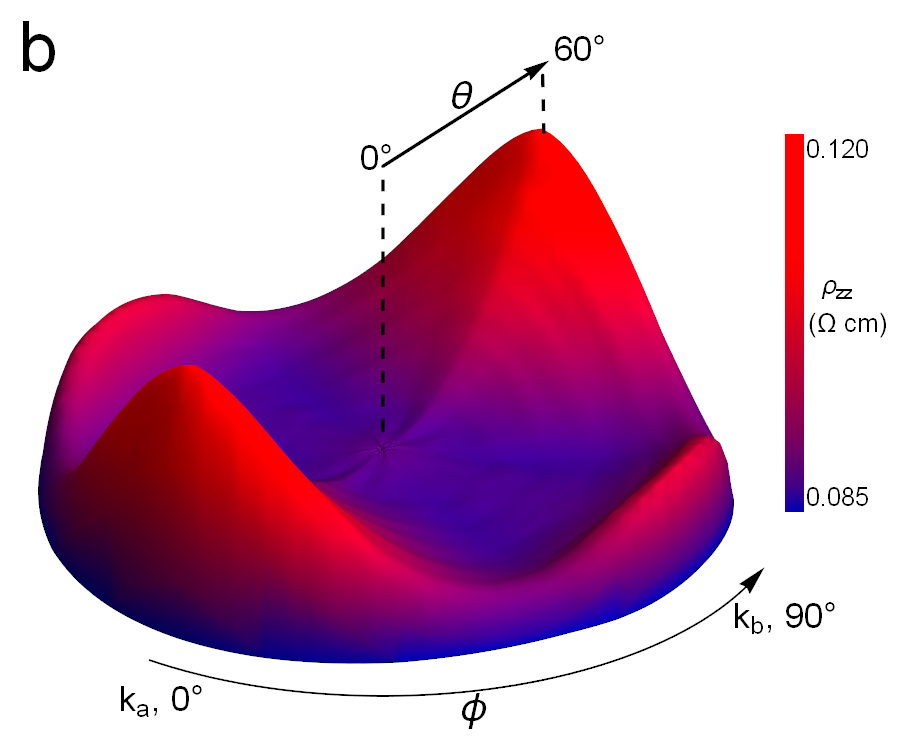}
\label{fig:osc3D}
}
\end{minipage}
\caption{\textbf{The angle-dependent magnetoresistance of \YBCO{58}.} The raw resistance (\textbf{a}) as a function of $\theta$ at 45 T and 15 K, for several values of $\phi$ spanning between the \aaxis ($\phi = \degr{0}$) and \baxis ($\phi = \degr{90}$). The rapid drop in resistivity beyond $\theta = \degr{60}$ is due to the onset of superconductivity when insufficient magnetic field is parallel to the \caxis. The inset defines the field angles $\phi$ and $\theta$ with respect to the crystallographic $\hat{\mathrm{a}}$, $\hat{\mathrm{b}}$, and $\hat{\mathrm{c}}$ axes. This data was taken at a temperature of 15~K to increase the polar angular range over which the normal resistive state is accessed \cite{Ramshaw:2012b}, and to thermally suppress QOs. Broken \Cfour symmetry is clearly shown in a polar plot of the data (\textbf{b}), where the radius is the polar angle $\theta$, and the amplitude and colour correspond to the magnitude of the resistance.}
\label{fig:osc}
\end{figure}

In order to understand the connection between AMR and the Fermi surface, we first perform a qualitative analysis of the three most salient features in the data: \Ctwo symmetry, negative AMR at low $\theta$ near the $\phi = \degr{90}$ directions, and suppression of AMR along the $\phi = \degr{45}$ direction at high $\theta$. The semi-classical conductivity of a metal is given by the velocity-velocity correlation function of all quasiparticles on the Fermi surface, averaged over their lifetime $\tau$ \cite{Chambers:1952,Abrikosov:1957}. Magnetoresistance arises because the quasiparticle velocity---which is always perpendicular to the Fermi surface---is altered by the Lorentz force which induces cyclotron motion perpendicular to the magnetic field (see Fig. 2a for a schematic cyclotron orbit). If $\tau$ is sufficiently long, orbiting quasiparticles sample a significant portion of the Fermi surface perimeter before scattering. Depending on the specific geometry of the cyclotron orbit, components of the quasiparticle velocity ($v_\mathrm{z}$, for example) may average to zero, resulting in a vanishingly small contribution to the conductivity for that field direction. The total magnetoresistance at a particular field orientation is then given by the ensemble of all quasiparticle orbits on the Fermi surface at that angle, and the AMR therefore encodes the Fermi surface geometry.
\begin{figure}
\begin{minipage}[]{.32\columnwidth} 
\subfloat{
\includegraphics[width=1\columnwidth]{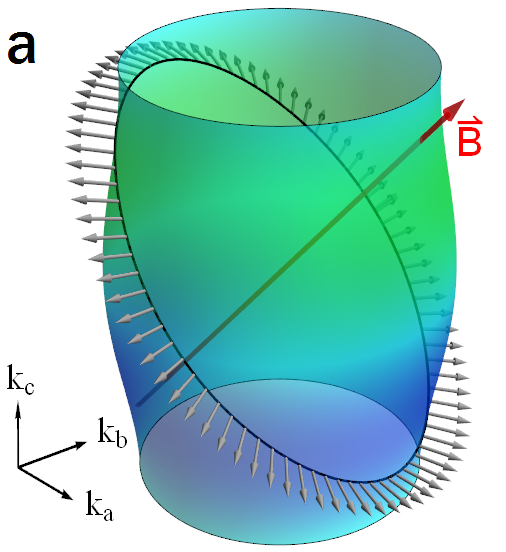}
\label{fig:orbit1}
}
\end{minipage}
\begin{minipage}[]{.65\columnwidth} 
\subfloat{
\includegraphics[width=1\columnwidth]{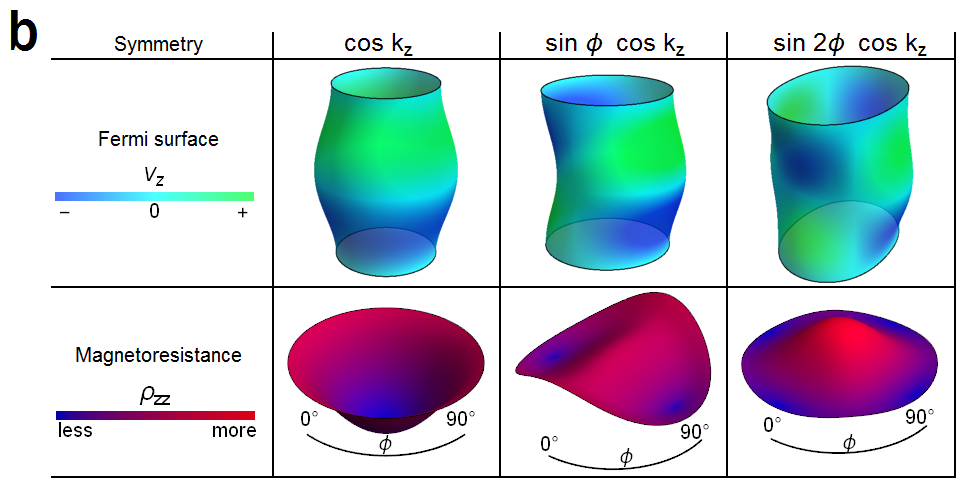}
\label{fig:orbits}
}
\end{minipage}
\caption{\textbf{Quasiparticle trajectory and magnetoresistance for different interlayer tunnelling symmetries.} Schematic Fermi surface with a cyclotron orbit (\textbf{a}). Quasiparticles experience a Lorentz force in a magnetic field (red arrow), inducing cyclotron motion perpendicular to the field direction. Grey arrows indicate the Fermi velocity along a cyclotron orbit (black line). Dispersion along the $k_\mathrm{z}$ direction modulates the $\hat{\mathrm{z}}$ component of the velocity and changes the sign of $v_\mathrm{z}$ around the cyclotron orbit. Panel (\textbf{b}) shows the three dominant \caxis dispersion shapes allowed by symmetry in \YBCO{58} (top panels) and their subsequent contributions to the AMR (bottom panels). Full rotational symmetry is preserved for the $\cos k_\mathrm{z}$ dispersion, while $\sin \phi \cos k_\mathrm{z}$ has \Ctwo symmetry. $\sin 2 \phi\cos k_\mathrm{z}$ has \Ctwo symmetry but produces \Cfour symmetric AMR once all cyclotron orbits are accounted for. }
\label{fig:orbit}
\end{figure}

Underdoped \YBCOd contains, at a minimum, two sections of Fermi surface due to its bilayer crystal structure (Fig. 3a). The phenomenon of magnetic breakdown, whereby quasiparticles in a magnetic field can jump between Fermi surfaces separated by small energy gaps \cite{Shoenberg:1984}, leads to multiple possible cyclotron orbits for the reconstructed Fermi surface of \YBCO{58} \cite{Sebastian:2014,Maharaj:2015,Forgan:2015}. For the purpose of analysing our AMR, we treat each possible `breakdown' path as a separate Fermi surface, with the surfaces contributing to the total conductivity in parallel. In general, the interlayer dispersion of a quasi-2D Fermi surface---which determines the interlayer resistivity \rhozz---can be expanded in cylindrical harmonics, three of which are shown in Fig. 2b. The simplest harmonic $\cos k_\mathrm{z}$ produces magnetoresistance that increases with $\theta$ for all $\phi$. The harmonic $\sin 2 \phi\cos k_\mathrm{z}$ produces four-fold symmetric AMR that decreases with increasing $\theta$, with a weaker effect along the diagonal directions. The simplest harmonic to break \Cfour symmetry in the interlayer tunnelling is $\sin \phi\cos k_\mathrm{z}$, which has two-fold symmetry and gives AMR that decreases with increasing $\theta$ along one $\phi$ direction and very weak AMR along the perpendicular direction. The measured twofold anisotropy of \rhozz in \YBCO{58}, along with the negative AMR at low $\theta$ near the $\phi = \degr{90}$ directions, therefore suggests that there is a significant contribution from a section of Fermi surface with $\sin \phi\cos k_\mathrm{z}$ symmetry. The most obvious that such a Fermi surface can exist by symmetry in \YBCO{58} is if the Fermi surface reconstruction itself breaks \Cfour symmetry---apart from a small orthorhombic distortion (and the copper-oxide chain layer whose contribution we rule out below), the unreconstructed Fermi surface of \YBCOd is nearly \Cfour symmetric \cite{Fournier:2010}. Our simulations suggest that anisotropy in the scattering rate; anisotropy in the interlayer hopping that is still finite in all directions; or an in-plane anisotropy in the Fermi wavevector (i.e. a Fermi surface elongated along one in-plane direction), cannot produce an effect of the magnitude we observe (see S.I. for details). 

Another prominent feature in the data is the suppression of AMR along the $\phi = \degr{45}$ direction, particularly above $\theta = \degr{50}$. In addition to the interlayer velocity, AMR is responsive to the in-plane geometry of the Fermi surface \cite{Singleton:2000,Bergemann:2003,Hussey:2003,Goddard:2004,Kartsovnik:2004,Analytis:2007}. For a cylindrical surface with simple $\cos k_\mathrm{z}$ warping and an isotropic Fermi radius \kF (see Fig. 2b), the AMR evolves with field angle $\theta$ as $\rhozz\left(\theta\right) \propto 1/\left(J_0\!\left(\kF c \tan \theta\right)\right)^2$, where $c$ is the \caxis lattice constant (see footnote \footnote{For more complicated warping geometries the actual form of $\rhozz\left(\theta\right)$ is different, but it is still the product $\kF c$ that sets the angular scale over which the maxima in \rhozz appear}). The AMR shows maxima wherever $\kF c \tan\left(\theta\right)$ equals a zero of the Bessel function $J_0$, which happens for certain `Yamaji' angles where the interlayer velocity averages to zero around the cyclotron orbit \cite{Yamaji:1989}. A critical feature of this form of \rhozz is that the product $\kF c$ sets the scale in $\theta$ over which these maxima in \rhozz appear: a smaller \kF pushes the resistance maxima out to higher angles. For quasi-2D Fermi surfaces that are non-circular in cross section (anisotropic $\kF\!\left(\phi\right)$), $\kF c$ still sets the angular scale, but to a first approximation it is $\kF\!\left(\phi_0\right)$---where $\phi_0$ is the direction in which the magnetic field is being rotated---that determines where $\theta$ and $\phi$ the maxima in \rhozz appear. For \YBCO{58} we can use the average value of $\kF \approx 1.27~\mathrm{nm}^{-1}$ determined by QOs to estimate that the first maximum in \rhozz should appear around $\theta \approx \degr{58}$. While this means that we cannot reach any AMR maxima before the onset of superconductivity, we are still able to observe the approach to the first maximum. The relatively weak AMR observed when the field is rotated in the $\phi = \degr{45}$ direction suggests that the Fermi surface has a smaller \kF along the diagonals than along the $\hat{\mathrm{a}}$ and $\hat{\mathrm{b}}$ axes---that its first maximum is pushed to higher $\theta$. This points to a Fermi surface with a square or diamond-like in-plane geometry---strikingly similar to the Fermi surface reconstruction predicted to occur via CDW \cite{Harrison:2011,Harrison:2012,Sebastian:2014,Maharaj:2014,Allais:2015,Briffa:2015}. Two scenarios that could give rise to this type of reconstruction are shown in Fig. 3.
\begin{figure}
\begin{minipage}[]{.35\columnwidth} 
\subfloat{
\includegraphics[width=1\columnwidth]{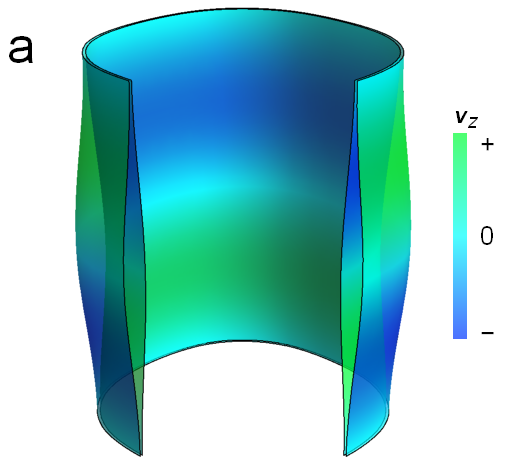}
\label{fig:recon0}
}
\end{minipage}
\begin{minipage}[]{.58\columnwidth} 
\subfloat{
\includegraphics[width=1\columnwidth]{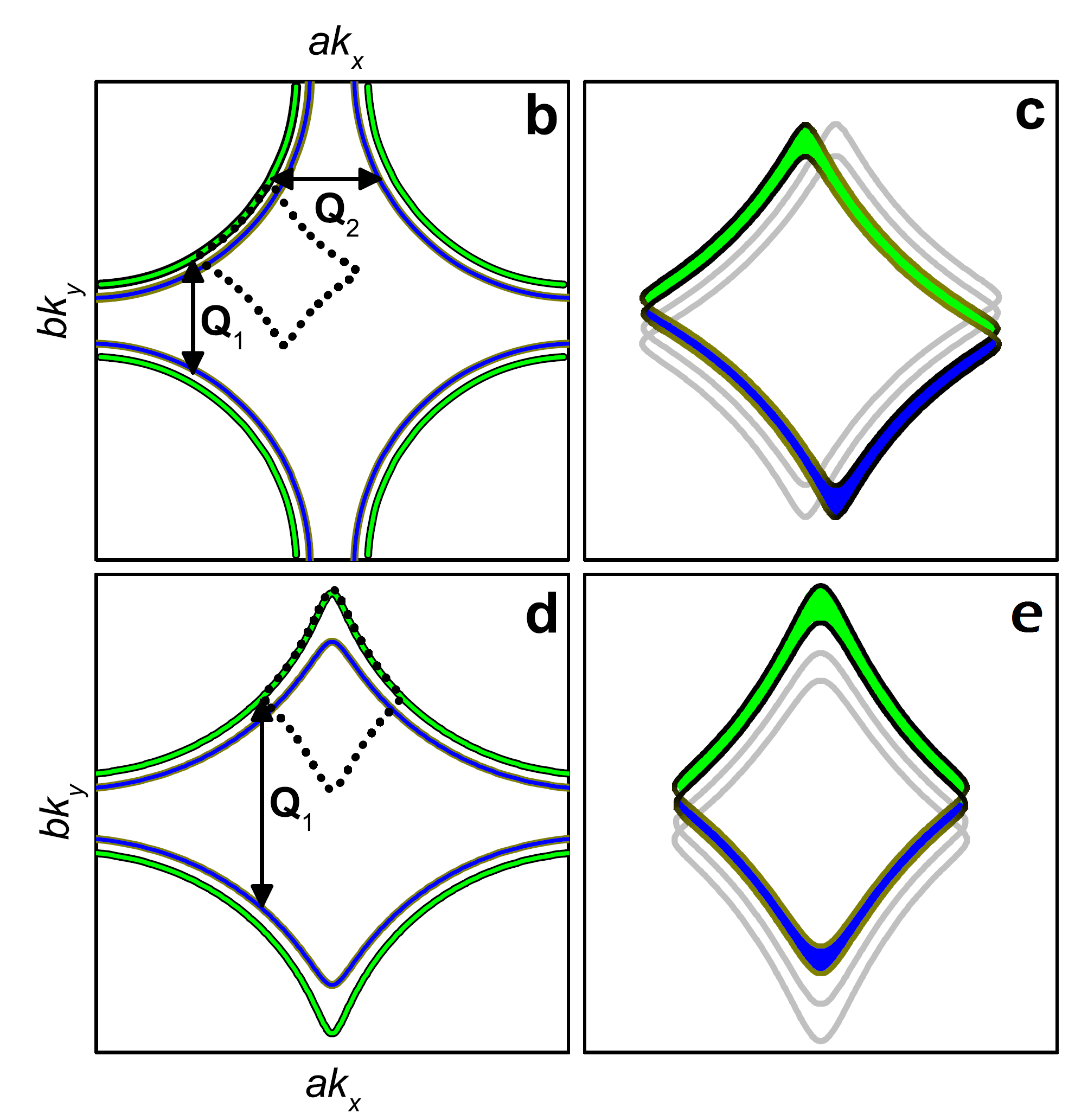}
\label{fig:recon1}
}
\end{minipage}
\caption{\textbf{Two possible anisotropic CDW reconstruction scenarios for \YBCO{58}.} The unreconstructed cuprate Fermi surface is a large hole-like cylinder \cite{Peets:2007,Hossain:2008,Hussey:2003} (\textbf{a}). The bilayer copper oxide planes of \YBCOd give rise to bonding and anti-bonding bands, whose interlayer velocities have opposite signs (one quarter of the Fermi surface has been cut away for clarity). The right hand panels are projections of this surface into the $k_\mathrm{a}-k_\mathrm{b}$ plane, from $k_\mathrm{z} = -\pi/c$ to $k_\mathrm{z} = 0$, before (\textbf{b} and \textbf{d}) and after (\textbf{c} and \textbf{e}) Fermi surface reconstruction. The colour scale signifies the sign of $v_\mathrm{z}$. The upper two panels involve two anisotropic CDWs, with $Q_2 = (0.31,0,1/2)$ and $Q_1 = (0.32,0,0)$ \cite{Blanco:2014,Gerber:2015,Chang:2015}, whereas the bottom two panels are the result of a large nematic distortion\cite{Yao:2011} followed by uni-axial CDW reconstruction with $Q_1 = (0.32,0,0)$. Multiple Fermi surface sections, between which magnetic breakdown is possible \cite{Harrison:2012,Sebastian:2014,Maharaj:2015,Forgan:2015}, are present after reconstruction: for clarity we colour only one section. We also assume that other sections of Fermi surface produced by these reconstructions are removed by the pseudogap. 
}
\label{fig:recon}
\end{figure}

We can now combine our qualitative Fermi surface information---interlayer tunnelling with \Ctwo symmetry on at least one Fermi surface section, and an in-plane diamond shape---and quantitatively model our data by numerically solving the Boltzmann transport equation \cite{Chambers:1952} (see S.I. for details). To avoid over-parametrization of the data we model only the three $F = 530$~T `breakdown' surfaces (plus their symmetry-related copies), which are known to dominate the \caxis conductivity \cite{Ramshaw:2011,Harrison:2015}. We fix the cross-sectional area of the orbits to the value of $A_k = 2\pi e/\hbar\cdot530~\mathrm{T} \approx 5.1~\mathrm{nm}^{-2} $ obtained from QO measurements on \YBCO{58} \cite{Ramshaw:2011}, and allow a single value of $\tau$ to vary as a free parameter for all three surfaces. We find that the AMR is best modelled by a sum of $41\%$ of a surface with $\sin \phi\cos k_\mathrm{z}$ symmetry, $27\%$ of a surface with $\cos k_\mathrm{z}$ symmetry, and $32\%$ of a surface with $\sin 2 \phi\cos k_\mathrm{z}$ symmetry. These proportions are reasonably in line with what is expected from the number of possible breakdown orbits and their magnetic breakdown probabilities \cite{Sebastian:2014}. In Fig. 4 we show that the most important features of the AMR which we first identified on qualitative grounds---\Ctwo symmetry, negative AMR near $\phi = \degr{90}$, and suppression along the ($\phi=$~45$^\circ$) direction at high $\theta$---are captured by this model. It is important to note that resistivity is not a linear function of the interlayer hopping parameters: combining all three symmetry warpings onto a single section of Fermi surface cannot reproduce the data. 

\begin{figure}
\begin{minipage}[]{.65\columnwidth} 
\subfloat{
\includegraphics[width=1\columnwidth]{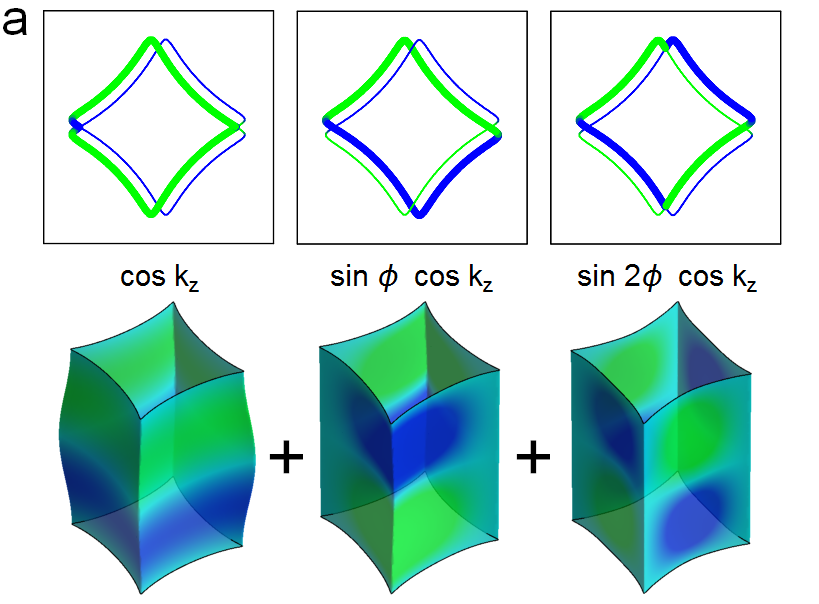}
\label{fig:surfs}
}
\end{minipage}\\
\begin{minipage}[]{.85\columnwidth} 
\subfloat{
\includegraphics[width=1\columnwidth]{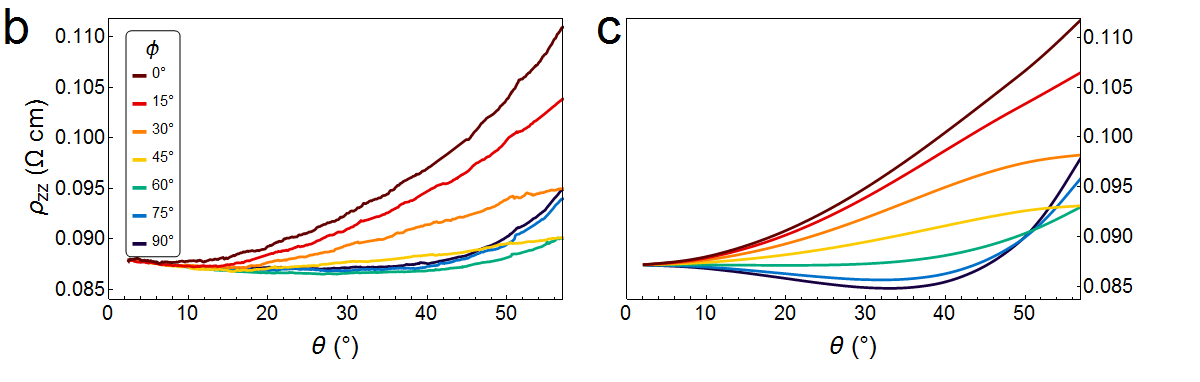}
\label{fig:sim1}
}
\end{minipage}
\caption{\textbf{Simulated magnetoresistance from a single bilayer-split pocket in \YBCO{58}.} The model Fermi surface consists of three dominant breakdown sections (\textbf{a}) that combine the in-plane diamond geometry of \autoref{fig:recon} with the interlayer dispersion symmetries of \autoref{fig:orbits}. The raw magnetoresistivity data is reproduced in panel (\textbf{b}), truncated at $\theta = \degr{58}$ at the onset of superconductivity. Panel (\textbf{c}) is a simulation of this data using the three magnetic breakdown orbits contributing to the dominant QO frequency, which have symmetries shown in \autoref{fig:orbits}. The strongest contribution comes from the cylinder with $\sin \phi\cos k_\mathrm{z}$ warping which explains the dominant \Ctwo symmetry we observe in the AMR.  }
\label{fig:sim}
\end{figure}

The consistency between our Fermi surface model and other experiments can now be checked. The quasiparticle lifetime we extract from the simulations is $\tau = 0.24 \pm 0.05$~ps---in agreement with the 0.27~ps reported from QOs \cite{Ramshaw:2012}. Zero-field resistivity measurements at this doping find an in-plane resistive anisotropy that is collapsing towards one at low temperatures: this behaviour is attributed to the conductivity of the 1D copper-oxide chains freezing out at low temperature \cite{Ando:2002}. Our model, which breaks \Cfour symmetry only in the \caxis dispersion of the Fermi surface, naturally preserves the near-isotropy of the \abplane conductivity. If the 1D copper oxide chain layer were clean enough to produce AMR---and there are is evidence from spectroscopy to believe that this is not the case \cite{Schlesinger:1990,Basov:1994,Bobowski:2010b}---then the resultant in-plane resistivity would be highly anisotropic: such in-plane anisotropy is not observed experimentally \cite{Ando:2002}. AMR from an open Fermi surface produced by the chain layer would also not produce the observed upturn in the AMR above $\theta = \degr{40}$ when $\phi = \degr{90}$ (see S.I. for details). AMR measurements can detect sections of Fermi surface that are invisible to QOs, such as open sheets and surfaces with higher scattering rates (shorter $\tau$). AMR therefore has the potential to observe sections of Fermi surface in \YBCO{58} formed by CDW reconstruction that have remained unobserved by QOs. Our simulation reveals, however, that the AMR can be fully accounted for by the same bilayer-split electron pocket that appears in QO experiments, with no additional sheets or pockets. This is consistent with previous suggestions that there is only a single Fermi pocket in the Brillouin zone of underdoped \YBCOd \cite{Riggs:2011,Sebastian:2011b} We note that the small hole pocket reported at this doping\cite{Doiron-Leyraud:2015} has a cross-section (\kF) that is too small to contribute any significant AMR below $\theta \approx \degr{75}$, thus we do not include the possibility of this surface in our model.

In order to obtain a Fermi surface with \Ctwo symmetry from the \Cfour symmetric unreconstructed Fermi surface \cite{Fournier:2010}, the mechanism of Fermi surface reconstruction must itself break \Cfour symmetry. We suggest that the diamond-like shape we measure with AMR strongly constrains the reconstruction scenario to one of density-wave origin. There is experimental evidence for broken \Cfour symmetry in the CDW: NMR measurements to 28.5 T are consistent with a high-field unidirectional CDW (`stripe') phase \cite{Wu:2011}, and recent X-ray measurements indicate that the CDW modulated along the \baxis---which has the same wavevector as the CDW modulated along the \aaxis in zero field---acquires a new \caxis component above 15 T\cite{Gerber:2015,Chang:2015}. We suggest that this asymmetry in the \caxis components of the wavevectors persists to at least 45 T, and that these CDWs are responsible for the Fermi surface reconstruction, as has been proposed theoretically \cite{Harrison:2011,Harrison:2012,Sebastian:2014,Maharaj:2014,Allais:2015,Briffa:2015}. A second route to a diamond-like pocket with \Ctwo symmetry in the \caxis tunnelling requires a strong nematic distortion of the Fermi surface (see Fig. 3 d and e)\cite{Yao:2011}. In high magnetic fields this surface is then reconstructed by a unidirectional CDW. This scenario only requires the single CDW wavevector that acquires three-dimensional coherence in high fields \cite{Gerber:2015,Chang:2015}. Independent of the specific model, our analysis shows that Fermi surface reconstruction in \YBCO{58} is anisotropic, and that the entire AMR signature can be accounted for by the same Fermi surface that is responsible for quantum oscillations (QOs) and a negative Hall coefficient in high magnetic fields. We have also established that although the CDW in the cuprates is relatively weak and disordered compared to more traditional CDW metals \cite{Andres:2003}, its symmetry and wavevector directly determine Fermi surface properties. This is similar to the iron pnictide superconductors, where the nematic state is directly observable on the Fermi surface \cite{Nakayama:2014}, and may suggest that the physics underlying both the superconductivity and the quantum criticality in these two classes of materials share a similar origin.

\section{Acknowledgments}
\label{se:Acknowledgments}
The authors would like to thank A. Damascelli and I. Elfimov for discussions on the bandstructure of \YBCOd. The authors would also like to thank S.A. Kivelson, S. Lederer, A.V. Maharaj, and L. Nie for helpful discussions. P.A.G. and S.G. thank the EPSRC for support. SES acknowledges support from the Royal Society, the Winton Programme, and the European Research Council under the European Union's Seventh Framework Programme (grant number FP/2007-2013)/ERC Grant Agreement number 337425.
This work was performed at the National High Magnetic Field Laboratory, which is supported by the National Science Foundation Cooperative Agreement No. DMR-1157490, the State of Florida, and the U.S. Department of Energy. N.H. and B.J.R acknowledge funding by the U.S. Department of Energy Office of Basic Energy Sciences ``Science at 100 T'' program.

\end{document}